\documentclass
[aps,floats,showpacs,prd,amssymb,floatfix,nofootinbib,balancelastpage,superscriptaddress,amsmath]{revtex4}

\usepackage{epsfig,epstopdf,float}

\begin{document}

\title{On the Dark Matter Solutions to the Cosmic Ray Lepton Puzzle}

\author{Pierre Brun}
  \address{CEA, Irfu, Service de Physique des Particules, \\ Centre de Saclay, F-91191 Gif sur Yvette, France}

\begin{abstract}
Recent measurements of cosmic ray leptons by PAMELA, ATIC, HESS and Fermi revealed interesting excesses. Many authors suggested particle Dark Matter (DM) annihilations could be at the origin of these effects. In this paper, we critically assess this interpretation by reviewing some results questioning the naturalness and robustness of such an interpretation. Natural values for the DM particle parameters lead to a poor leptons production so that models often require signal enhancement effects that we constrain here. Considering DM annihilations are likely to produce antiprotons as well, we use the PAMELA antiproton to proton ratio measurements to constrain a possible exotic contribution. We also consider the possibility of an enhancement due to a nearby clump of DM. This scenario appears unlikely when compared to the state-of-the-art cosmological N-body simulations. We conclude that the bulk of the observed signals most likely has no link with DM and is rather a new, yet unconsidered source of background for searches in these channels.
\end{abstract}

\maketitle


\section{Cosmic ray electrons and positrons : measurements and conventional production}

\begin{figure}
\centering
\epsfig{figure=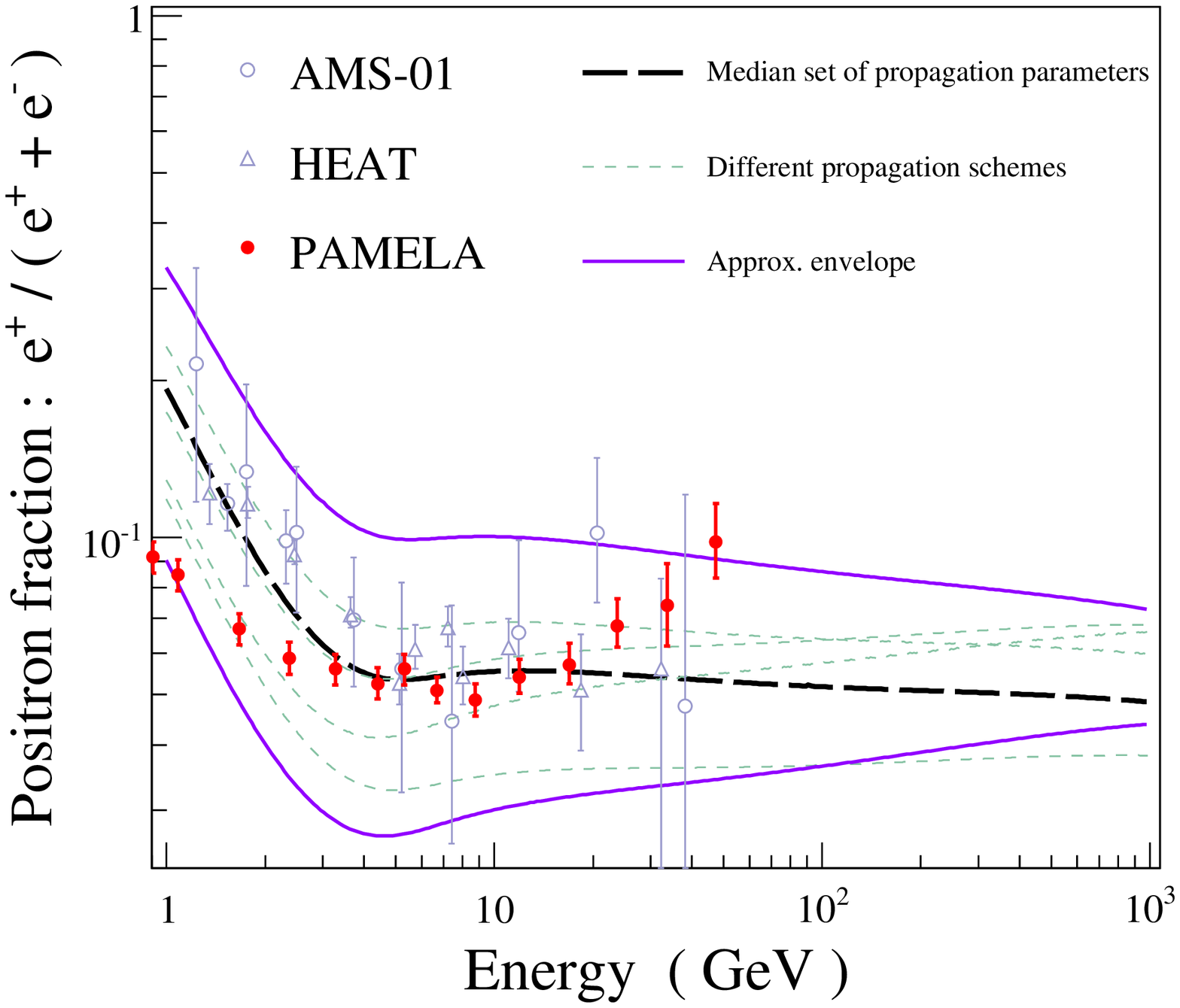,height=2.5in}
\epsfig{figure=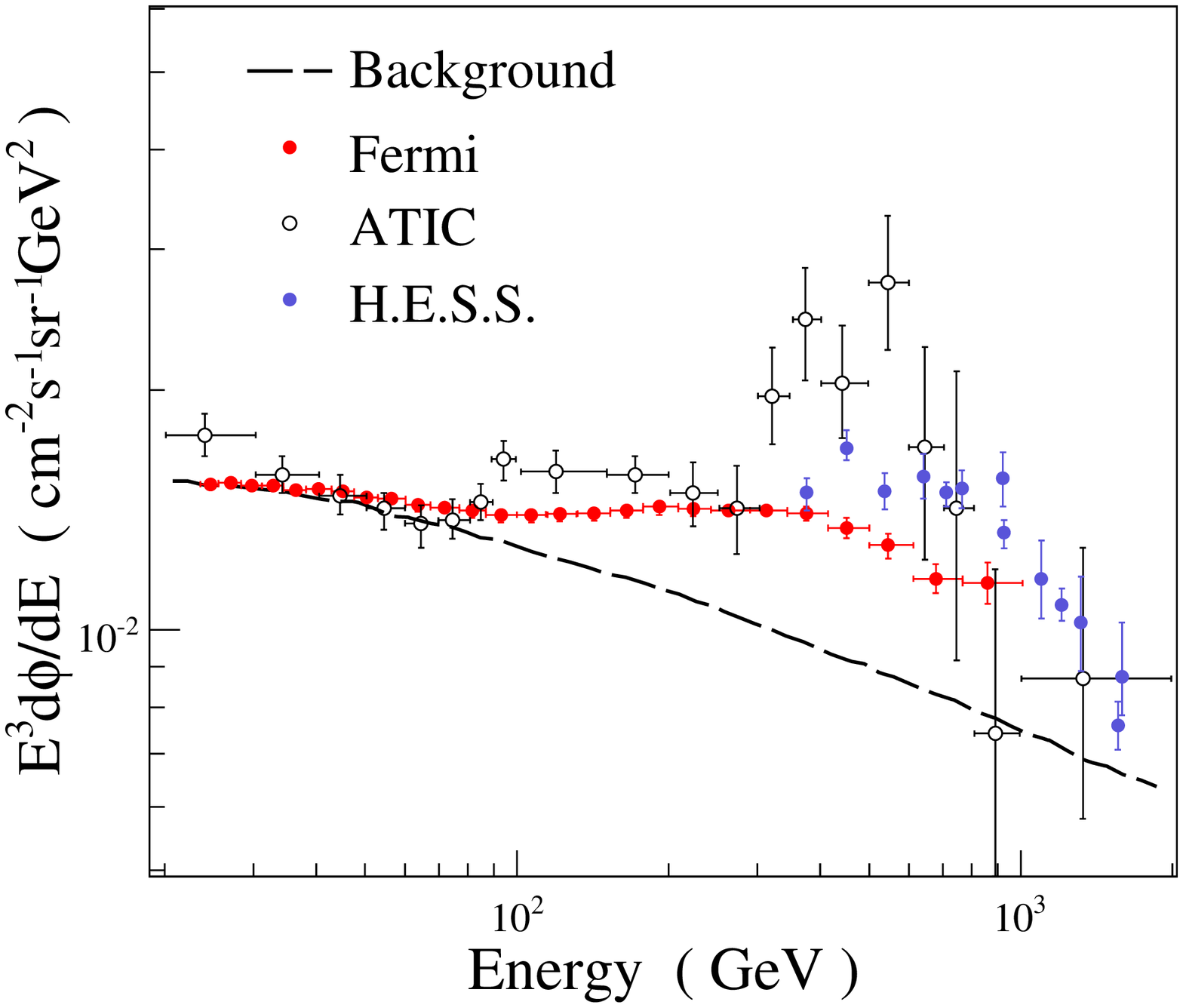,height=2.5in}
\caption{Cosmic ray positron fraction (left) and electrons+positrons fluxes (right).
\label{fig1}
}
\end{figure}

Recent measurements of cosmic ray positrons and electrons in the GeV-TeV region have rised a lot of interest as they exhibit unexpected features. The PAMELA satellite found evidence for a rise of the positron fraction ($e^+/(e^-+e^+)$) above 10 GeV~\cite{pamela}, as some previous experiments found some hints for~\cite{ams,heat}. The data points are shown on the left panel of Fig.~\ref{fig1}. In an independent way, the ATIC, HESS and Fermi collaborations reported measurements of the $e^-+e^+$ flux, between dozens of GeV and a few TeV~\cite{atic,hess,fermi}. These seem to indicate a deviation from a pure power law above 100 GeV, as shown on the right panel of Fig.~\ref{fig1}. Note that sizable systematic errors affect these data so that it is still unclear whether the feature is a sharp peak or a smoother bump.

In the standard picture, primary cosmic rays (mostly protons) are produced and accelerated in supernovae and their remnants, and secondaries result from the spallation of primaries onto the interstellar medium. As conventional sources carry the same baryon number as the whole Universe, electrons are mostly primaries and positrons are usually though to be secondaries. It is extremely difficult to reproduce the PAMELA rise or the leptonic sum feature in conventional diffusion models. As an illustration, the left panel of Fig.~\ref{fig1} displays a collection of predictions from a diffusion model. The different curves correspond to distinct sets of parameters allowed by cosmic ray measurements. None of propagation setups can reproduce the data (for more details on diffusion models, see~\cite{delahaye} and references therein). It is yet unclear if the $e^-+e^+$ feature and the positron rise have a common origin, but they both seem to indicate the presence of a nearby primary source of cosmic ray electrons and positrons (leptons), which is not accounted for in conventional cosmic ray models. Due to the efficient energy losses of the electrons and positrons as they propagate in the Galaxy, this primary production must take place within a short $\sim$1 kpc radius around the Earth. 

Dealing with nearby primary sources of positrons, one can generally invoke two classes of models: pure astrophysics and particle physics inspired models. In the first case, it is assumed that some astrophysical source exists in our Galactic neighbourhood which produces an abnormal amount of positrons. This could be for instance a supernova remnant with high level of secondary particles produced at the source, or a pulsar. It is actually quite easy to reproduce the leptonic signals with conventional sources but there is still no obvious candidate. Particle physics inspired interpretations of the data relying on new physics beyond the Standard Model, the second possibility is more speculative. Most studied possibilities consider DM annihilation or decays as exotic source of cosmic rays. For a review of the measurements and possible interpretations, from both astrophysics and particle physics, see~\cite{tango}. In this paper, we focus on the annihilating DM case, we question in particular the naturalness of such an interpretation. As some conventional interpretations for the cosmic ray leptons excesses exist, it can seem odd to invoke more exotic scenarios. This is however an extensively studied possibility and it is now important to study how natural and robust the DM hypothesis can be.

\section{Dark matter annihilations as a source of cosmic rays}

\subsection{Canonical models and enhancement of the signal}

One of the main motivations for particle DM in the form of Weakly Interacting Massive Particles (WIMPs) relies on the estimation of their cosmological relic density. In the canonical models, this density is regulated by primordial self-annihilations, which rate is proportional to the annihilation cross section. An electroweak-scale value for the cross section ($\langle \sigma v \rangle = 3\times 10^{-26}\;\rm cm^3 s^{-1}$) leads to the correct relic abundance of $\Omega_{cdm}h^2\simeq 0.1$. This numerical coincidence is often referred to as the `WIMP miracle'. In our Galactic halo, DM annihilations can still take place, with a rate at the location $\vec{x}$ of 
\begin{equation}
\Gamma(\vec{x})\;\propto\;\rho^2(\vec{x})\times\frac{\langle \sigma v \rangle}{m^2}\;\;,
\label{eq1}
\end{equation}
$\rho$ being the DM density and $m$ the WIMP mass. Annihilations of DM particles produce standard particles and antiparticles in a 1:1 ratio, thus enlarging the antimatter fraction. Produced particles can be directly lepton pairs or other particles like quarks, the hadonization of which will produce electrons and positrons. As a general rule, the more direct the production in the annihilation process, the harder injected spectrum. The PAMELA rise favors a hard injection spectrum, it could be for instance direct production $\chi\chi\rightarrow e^+e^-$ or $\chi\chi\rightarrow W^+W^-$. These exotically produced leptons would then propagate in the Galaxy before reaching our detectors. Doing so, the spectral information on the production channel gets somehow erased.

Though very seducing, the DM interpretation implies a difficulty. In fact, the canonical model presented above leads to a signal which is 2 to 3 orders of magnitude too small to produce the observed effects. There are several motivated reasons why the signal could be enhanced, which fall into two types whether one considers an increase linked to $\langle \sigma v \rangle$ (particle physics type) or to the DM distribution $\rho$ (astrophysics inspired type). In the following, enhancement has to be understood as enhancement with respect to this canonical values.

\begin{figure}
\centering
\epsfig{figure=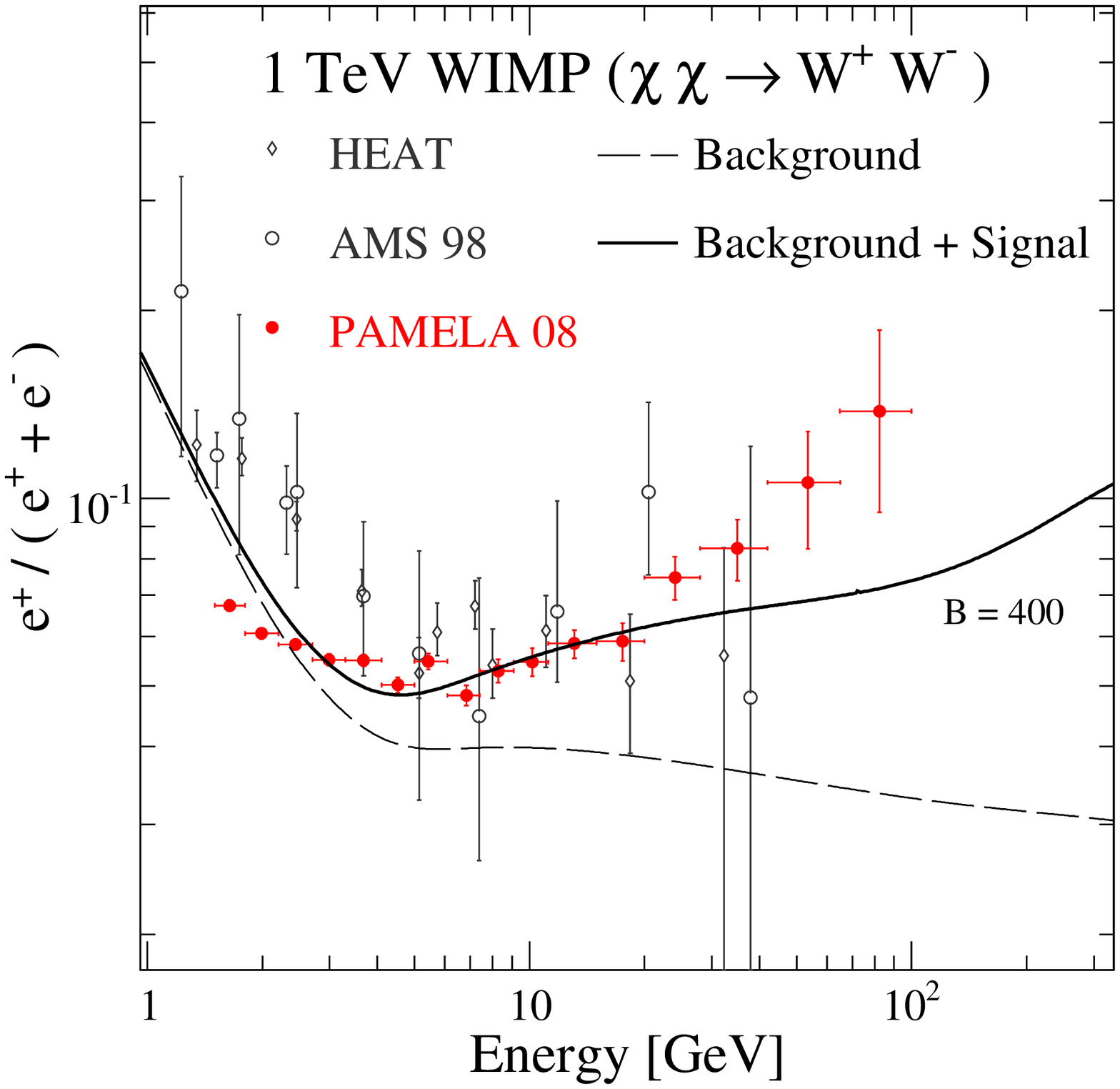,height=2.7in}
\epsfig{figure=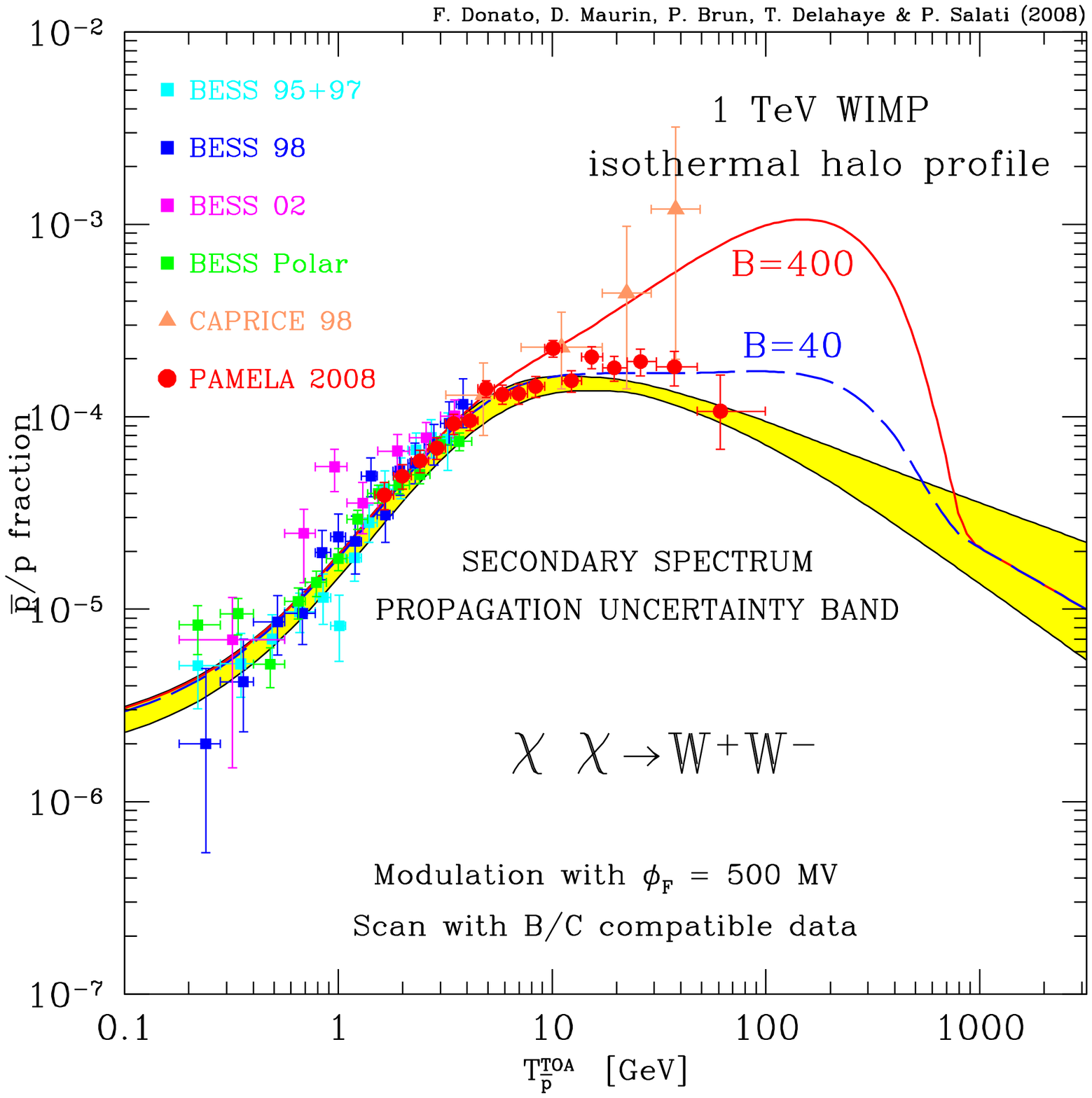,height=2.7in}
\caption{ A model reproducing the positron fraction rise (left panel), but overproducing antiprotons (right panel).
\label{fig2}}
\end{figure}

In the first case, some effects can lead to a substantial increase of the cross section when TeV-scale mass WIMPs are considered, due to resonances ({\it e.g.} the Sommerfeld effect), this case is discussed in the present section. The second case where the signal is increased by the effect of DM distribution is addressed in Sec.~\ref{clumps}. On the left hand-side of Fig.~\ref{fig2} is displayed a possible interpretation of the PAMELA data with 1 TeV WIMPs annihilating into gauge boson pairs. In that case, the positron fraction rise can be reproduced, to the price of an increase of $\langle \sigma v \rangle$ by a factor of 400. With the help of the Sommerfeld effect, a case for such an enhancement of the natural cross-section can easily be made from the model-building perspective. The Sommerfeld enhancement appears in case an attractive interaction ($e.g.$ through conventional electroweak bosons) can lead to almost-bound states of DM particles. Because of that, the slower the particles, the higher the cross section. This mechanism allows to have a high annihilation cross section in the (nowadays cold) DM halo with an annihilation rate in the early Universe still low enough to get the correct relic abundance. A drawback of these models is that the DM particle mass must be close to some resonance for the amplification to be sizeable. Then a substantial level of fine tuning is required.

\subsection{Tightening the net : constraints from other messengers}

\begin{figure}[h]
\centering
\epsfig{figure=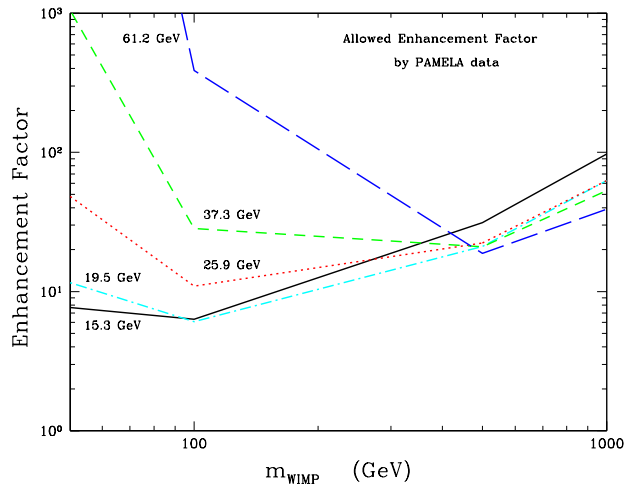,height=2.5in}
\caption{Maximum enhancement factors allowed by different PAMELA $\bar{p}/p$ data points.
\label{fig_BF}}
\end{figure}

In this section, we explore the implications of given interpretations of the leptonic data for other observables. This helps to reduce the possibilities regarding the particle dark matter configurations. In case the leptons are not directly produced in the annihilation process but rather result from the hadronization of colored particles, antiprotons should be produced as well. The PAMELA satellite also precisely measured the antiproton to proton ratio up to 100 GeV. In the right panel of Fig.~\ref{fig2}, data points are displayed together with theoretical expectations for the antiprotons conventional flux. It is noticeable that --contrary to the positron case-- our comic ray model nicely reproduces the antiproton data. It might be an indication that we are missing something in the case of leptons. As discussed previously, this something could be a primary source of electrons and positrons. The semi-analytical model used to preform these predictions allows to compute the theoretical uncertainty  for this prediction, represented here with a yellow band. To obtain it, a scan over all propagation parameters allowed by other cosmic ray observables is performed. Note that here the lowest possible value for the proton flux is used. It is possible then to have a slightly higher $\bar{p}/p$ ratio that fits the data better. This choice however leaves more room for a DM signal and thus allows to put more robust constraints on the enhancement factor. The obtained upper limits on the cross-section enhancement factor are of order $\sim$10 (see~\cite{donato}). An illustration is given in the right panel of Fig.~\ref{fig2}, which displays the antiproton counterpart for the same model as in the left panel. One can see that a boost factor of only 40 is permitted otherwise antiprotons are overproduced. The maximal allowed value for the enhancement can be computed from all single bins of the PAMELA measurement, and depends on the value of the DM particle mass. These extremal values are shown in Fig.~\ref{fig_BF}, maximum boost factors are $\sim$10 for DM masses below a few hundreds GeV and are allowed to reach up to $\sim 40$ if the DM particle is heavier.

As discussed above, an enhancement of $\sim$10 as allowed by antiprotons is not sufficient to reproduce the leptonic data. In order to save the DM interpretation of those, one has to assume that the hadronic content of the annihilation products is negligible, therefore producing no antiprotons. This would be precisely the case if DM particle annihilated into leptons only. This is the so-called leptophilic DM. 

In the leptophilic DM scenario, leptons are produced with a high efficiency in the vicinity of the Earth, producing the cosmic ray anomalies. In this framework, numerous leptons should also be produced in regions where DM in concentrated (Galactic center, dwarf galaxies, etc.). A closer look at the expected signals in $\gamma$-rays or radio shows that this scenario is hardly compatible with existing constraints~\cite{bertone}. For example, high energy leptons produced with a high rate in the region of the Galactic center should emit radio photons when being depleted by the magnetic field. Such a radio emission not being observed, one has to invoke non standard models for the magnetic fields in the region of the Galactic center.

Up to now, we saw that the production rate of DM-induced exotic leptons has to be increased with respect to canonical models. Obviously, the annihilation rate has to be higher than it was in the early Universe (otherwise there would be no more DM is the Universe). Looking back at equation~\ref{eq1}, two options are available : increasing either $\langle \sigma v \rangle$ or $\rho$ in the Earth neighborhood. In this section we investigated the first possibility and showed that we are forced to assume DM only annihilates into leptons. This scenario is in tension with independent measurements like for example radio data from the Galactic center. Then, it could be that the annihilation rate is high in the Earth neighborhood  but not in some other locations in the Milky Way. This could be a possible way to save the DM interpretation,  it corresponds to the increase of $\rho$. This can arise with the help of DM substructures, as discussed in the next section.

\subsection{Dark matter clumpiness}
\label{clumps}

In our Galaxy, the DM density is not expected to be smooth but rather partly lumpy. About 10\% of the halo mass is expected to lie in DM substructures, whose masses may range from Earth-mass to dwarf galaxies masses. An interesting possibility regarding the lepton anomalies would be that DM annihilations responsible for the excesses take place within a nearby subhalo. We explore here the possibility that a conventional clump is large and close enough to harbor efficient DM annihilation leading to the observed anomalies. In Ref.~\cite{brun}, we fit the properties of such a clump (distance, DM luminosity) to reproduce the cosmic ray data, with the assumption of a thermal relic density $\sigma v = 3\times 10^{-26}\;\rm cm^{3}s^{-1}$. Different annihilation channels are considered and it is found that leptophilic scenarios fit the data best, as expected. 

Note however that it has been previously shown that if large boost factors can exist is some extreme scenario, those are expected to be different for positrons and for antiprotons~\cite{brun0}. In our case, it is interesting to notice that up to 10\% of hadronic annihilation product can be accommodated in the case of a clump, in contrast with the case when the cross section is enhanced. The reason for that is the propagation of antiprotons which significantly differs from that of leptons. High energy leptons are found close to the source whereas antiprotons can propagate in the whole Galaxy without major energy losses. This dilution of the high energy signal for antiprotons implies that if the annihilation is located within a nearby subhalo, the enhancement factor for antiprotons is less than for leptons and consequently allows a larger hadronic production in the annihilations than in the case of a cross section enhancement.

Typical clumps that fit are $\sim$1 kpc away and present DM luminosity of order $10^8\;\rm M^2_{\odot}pc^{-3}$, the luminosity $L$ being the quantity
\begin{equation}
L\;=\;\int_V \rho^2 d^3\vec{x}\;\;.
\end{equation}

\begin{figure}[h]
\centering
\epsfig{figure=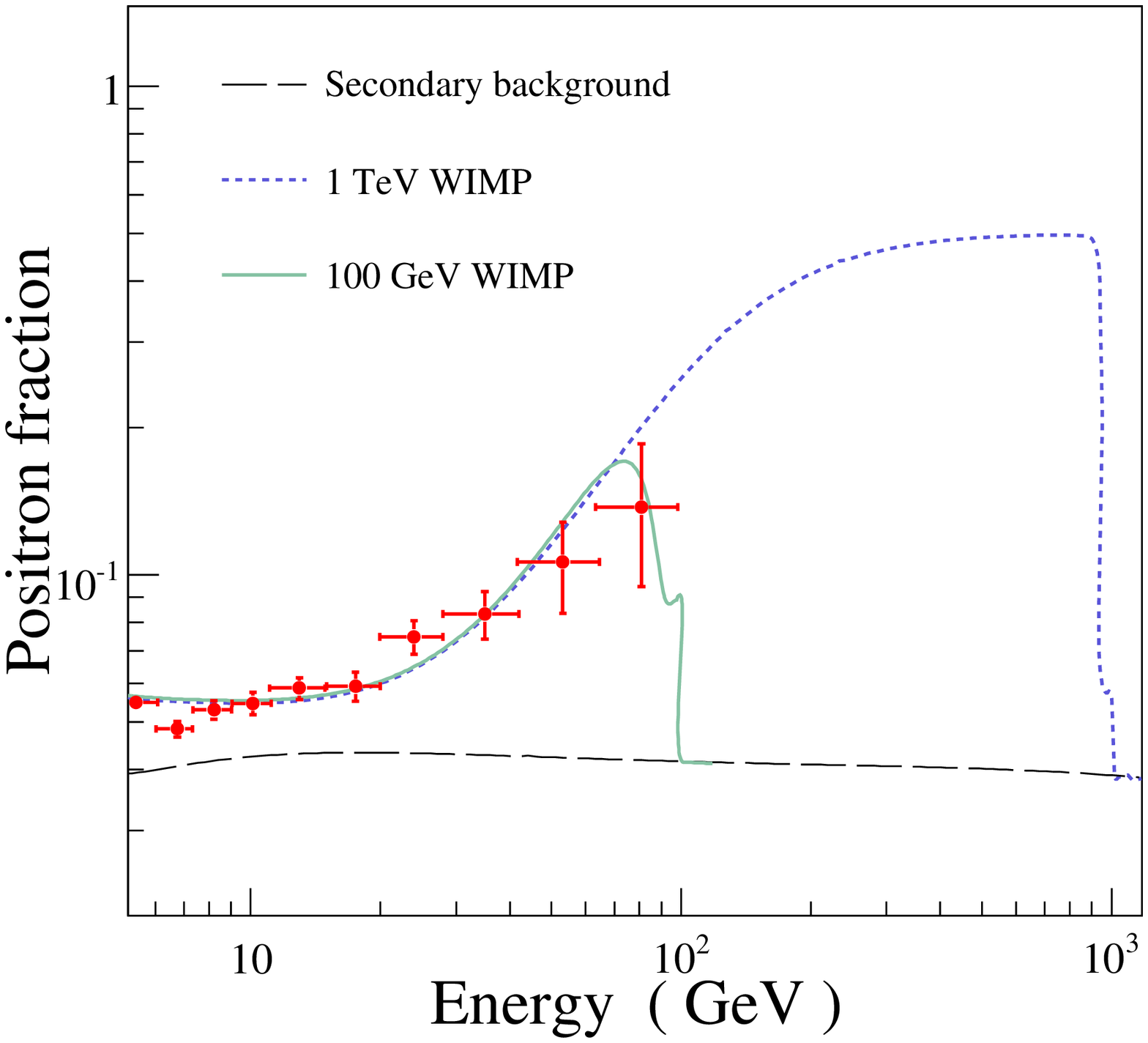,height=2.7in}
\epsfig{figure=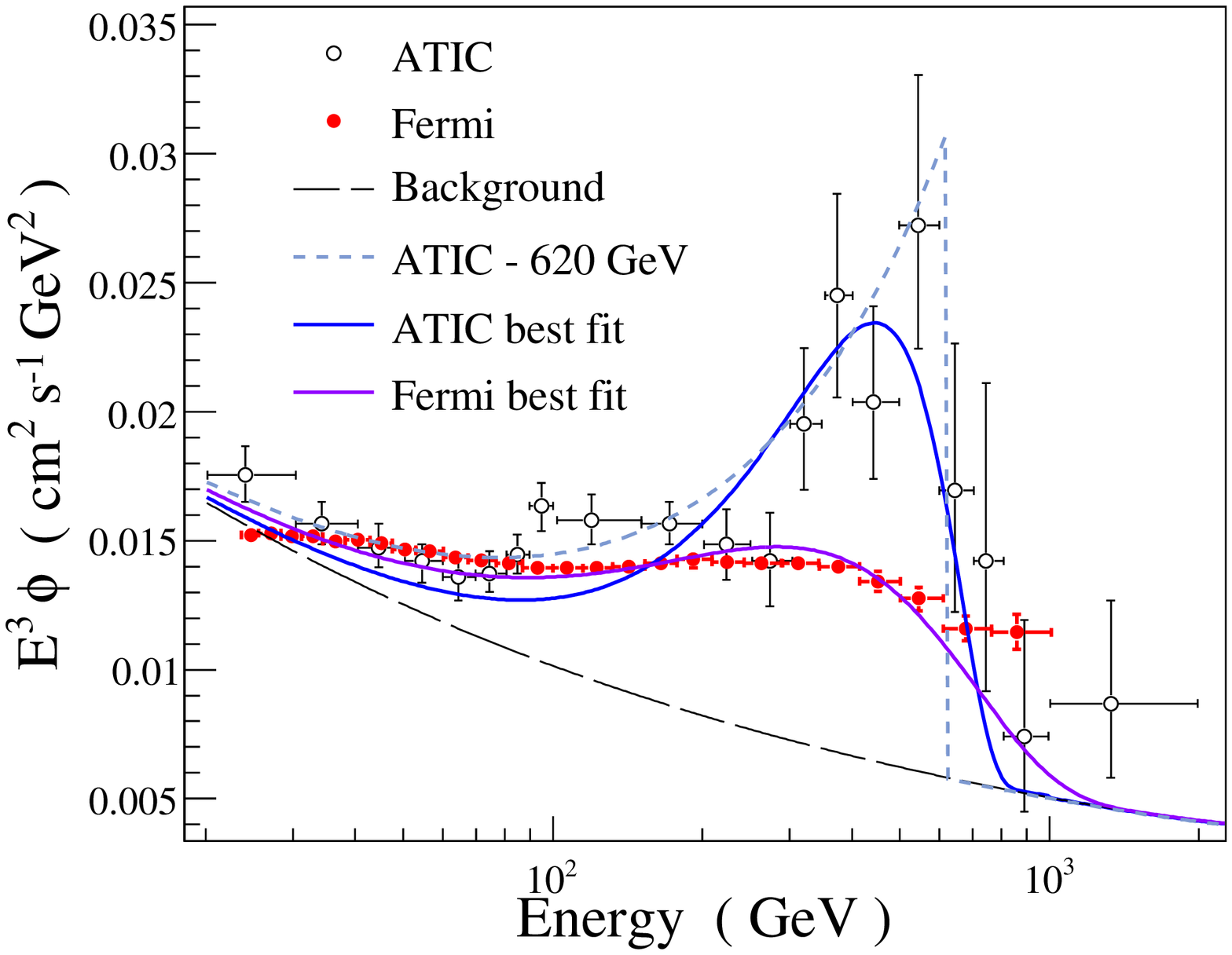,height=2.7in}
\caption{Best fits to the PAMELA data in the case of a positronic line (see the e$^{+}$/e$^{-}$ row of Tab.~\ref{tab1}) (left panel) 
and fits to the ATIC data (right panel).
\label{fig_fitclumps}}
\end{figure}

The complete properties (distance and luminosity) of clumps best fitting the data (fraction, sum or both) are detailled in Tab.~\ref{tab1}, for different annihilation channels and particle DM masses. The corresponding best-fits are displayed in Fig.~\ref{fig_fitclumps}. In the right panel of this figure, the case of a 620 GeV WIMP annihilating with a high cross-section (not within a clump) is shown for a matter of comparison, as this scenario is quoted in the original ATIC paper.

\begin{table}[t]
\centering
\begin{tabular}{|c||c|c||c||c|}
\hline
& \multicolumn{2}{|c||}{PAMELA} & ATIC & Fermi \\
\hline
m$_{\chi}$ (GeV) & 100 & 1 000  & 1 000 & 2500\\
\hline
\hline
e$^{+}$/e$^{-}$ &
$1.22 - 1.07 \!\cdot\! 10^7$ & $0.78 - 3.56 \!\cdot\! 10^9$ & $1.52 - 2.98 \!\cdot\! 10^9$ & $2.68 - 5.53 \!\cdot\! 10^{10}$\\
\hline
e$^{\pm}$ + $\mu^{\pm}$ + $\tau^{\pm}$ &
$0.44 - 2.51 \!\cdot\! 10^7$ & $0.27 - 9.84 \!\cdot\! 10^9$ & $0.25 - 8.78 \!\cdot\! 10^9$ & $2.81 - 2.17 \!\cdot\! 10^{11}$\\
\hline
\end{tabular}
\caption{Best fit values of the ($D ; L$) couple in units of
$(\rm{kpc} ; \rm{M}_{\odot}^{2} \, \rm{pc}^{-3})$
for various DM particle masses and annihilation channels.\label{tab1}}
\end{table}

The conclusion of Fig.~\ref{fig_fitclumps} is that very good fits of the data can be found with a single DM clump. Because of the characteristics of the electrons and positrons diffusion, we assume here that only one bright clump contributes to the local flux. This assumption is justified a posteriori in the following. Actually, it could be that two or more clumps contribute to building the signal but as we shall see, having one sufficiently bright clump is already unlikely.

\begin{figure}[h]
\centering
\epsfig{figure=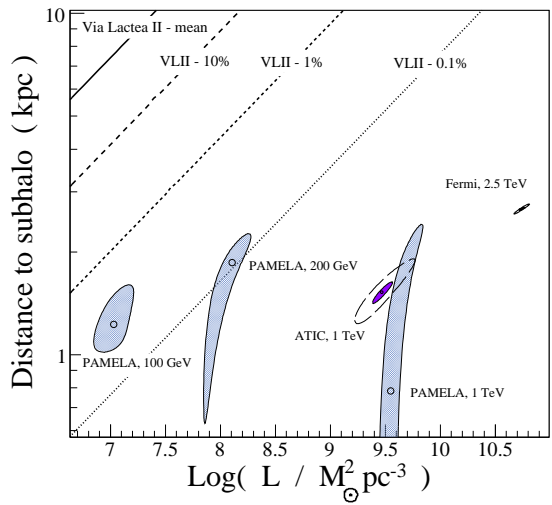,height=2.8in}
\caption{ Best fit results in a clump luminosity-distance plane for different configurations,
together with probabilities inferred from Via Lactea-II results.
\label{fig_proba}}
\end{figure}

Next step then is to evaluate how likely such clump configuration is. To do so, the state-of-the-art cosmological simulation of structure formation Via Lactea II~\cite{diemand} is used. Clumps that best fit the cosmic ray data are resolved in the simulation so that is it possible to explicitly compute the probability for a given clump to lie at a given distance. Fig.~\ref{fig_proba} shows the probabilities to appear for DM clumps in a luminosity-distance plane, inferred from Via Lactea II results. The oblique lines represent the equiprobability contours for Via Lactea subhalos in the distance-luminosity plane. This plot is very useful as it helps evaluating the natural distance at which a clump of a given luminosity lies. For instance, a subhalo with a luminosity of $7\;\rm{M}_{\odot}^{2} \, \rm{pc}^{-3}$ typically stands at a distance of $\sim$7 kpc from the Earth. In only 10\% of the Milky Way realizations such an object would be present at 4 kpc, and it appears at 2 kpc 1\% of the time. The different points show the best fit values for the clump parameters in different cases and the surrounding contours show the 1 $\sigma$ excursions around the best fits to the cosmic ray leptons data. It appears that the probability for a clump to reproduce the PAMELA data is only 0.37\%. Fitting simultaneously the ATIC peak is feasible, this is deduced from the fact that the contours from the different fits intersect. However in that case, the probability drops to $3\times 10^{-5}$. Finally, fitting the Fermi data is basically associated to a null probability. In that last case, one needs a subhalo which has almost the same luminosity as the whole Milky Way. Obvioulsy, the conclusion is that is it quite unlikely that the lepton excesses are caused by a nearby DM subhalo. We show here that it is possible, but with a very low probability. Should it be the case anyway, Fermi would detect the associated $\gamma$-ray emission within a few years of operation time.

As a remark, we point out that all the fitted spectra would have been obtained with significantly less luminous subhalos should the annihilation cross-section be directly enhanced. If the Sommerfeld effect  is at play, the signal from small clumps is enhanced with respect to the contribution from larger substructures since the velocity dispersion of DM particles decreases with the mass of the host subhalo. The blue contours of Fig.~\ref{fig_proba} are shifted towards smaller values of L and get nearer to the mean predictions of the VL-II simulation, with a much larger probability of occurrence. For instance by simply assuming $L \rightarrow L \times c / V_{max}$, the probability of ATIC best fit case increases from $p \sim 3 \times 10^{5}$ to 14 percent! This means that combining the effects of the different enhancement mechanisms can change the conclusions of this section. However, in that case instead of having one extreme scenario, we invoke two less extreme ones. It might be that a reasonable Sommerfeld enhancement is at stake within a more or less reasonable nearby clump. However with this scenario the question of fine tuning is back and such an explanation should probably be considered in case all other possible explanations --such as pulsars for example-- are ruled out.

\subsection{A remark about the energy dependence of clumpiness enhancement}

An important point is that for clumpiness enhancement, the boost factor depends a lot on the energy and lead to substantial spectral modulations. This was not the case for instance when increasing the cross section. In that case the annihilations in the very local smooth DM halo are sufficient, and the whole signal can basically be multiplied by an energy-independent number. A common  mistake is to treat clumpiness enhancement and increase the exotic flux without considering possible spectral modulations. Indeed, a specificity of lepton propagation is that regardless of the overall normalization, a feature observed at energy $E$ can always be produced by a source at distance $D$ with an injection energy $E_s$ as long as $ D^2 \propto E^{0.3} - E^{0.3}_s$ (the 0.3 exponent depends on the propagation setup). Therefore there is a degeneracy in the possible interpretation of the features. 

As an illustration, we show in Fig.~\label{fig_modulation} how the signal from DM can exhibit a double-bump feature if, in addition to the contribution from a smooth DM halo, two nearby clumps are taken into account (solid light-colored line). In this example, the subhalos lie at a distance of 0.9 and 4.3 kpc from the Earth and their luminosities are of order $10^8\;\rm M_{\odot}pc^{-3}$ and $10^{10}\;\rm M_{\odot}pc^{-3}$ respectively. Notice that, without additional enhancements (from $\langle \sigma v\rangle$), having such bright nearby clumps is practically ruled out and we use them here for pedagogical purpose.  As regards the dased curve of the right panel of Fig.~\ref{fig_modulation}, the sharp edge at 800 GeV is associated to a strong local DM annihilation and is produced by the VL-II smooth halo whereas the bump at $\sim$100 GeV comes from a single nearby clump located at 3.2 kpc. The cross section has been increased up to a value of $10^{-23}\;\rm cm^3s^{-1}$. This case seems somewhat more probable than the 2-clumps configuration. These examples illustrate how tricky boost factors are. Shifting upwards the DM cosmic ray fluxes turns out to be wrong especially in the light of the subbtle effects introduced by propagation. In the light of these results, it appears hopeless to deduce any information about the injection spectrum (such as the annihilation channel) from the measured spectral shape.

\begin{figure}
\centering
\epsfig{figure=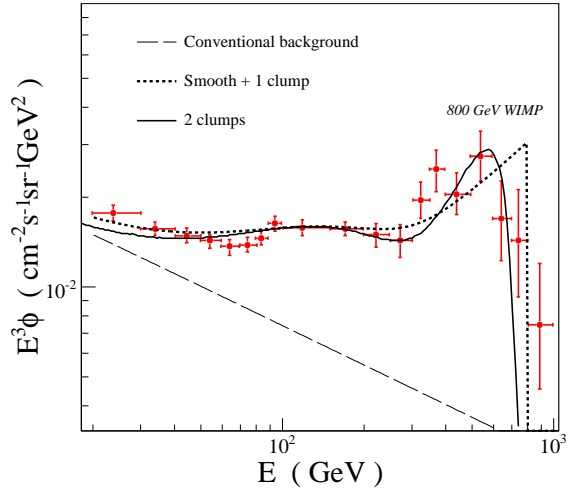,height=2.8in}
\caption{Electron-positron spectra resulting from one (dotted) or two (solid) nearby subhalos. In the latter case, a 620 GeV
DM species with thermal annihilation cross-section is considered.
\label{fig_modulation}}
\end{figure}

\section{Conclusions}

In this paper, it is shown that most likely the lepton excesses observed by PAMELA, ATIC, Fermi and HESS are not caused by conventional secondary cosmic rays, meaning that a nearby source significantly contributes to the local flux. The main classes of candidate are astrophysical ($e.g.$ a nearby pulsar) or more exotic (DM annihilations). The first case is clearly favored as one has to invoke non conventional scenarios for DM in order to avoid contraints from different observables. In conclusion, though very exciting, the DM interpretation of the cosmic ray excesses lacks naturalness, and a more conventional source might be at the origin of this phenomenon.

It is definitely possible to reproduce the observed cosmic ray data with the help of dark matter signals. Within this hypothesis however, there will always be some tension between the different channels and observables, or quite a high level of fine-tuning. It could be that we are encircling the dark matter particle properties but most likely, the bulk of the observed leptons come from a nearby astrophysical source producing a large fraction of electron/positron pairs. In that case, it would constitute an additional background for indirect searches for dark matter through lepton channels that had never been accounted for before. A big step forward will be the measurement of the small anisotropy in the arrival directions of the cosmic ray leptons, if any. If it is observed
and points towards a known pulsar, the conclusion will be quite clear, whereas if it points at some Fermi unidentified source outside the Galactic disk, dark matter might be back in the game. It also urges to separate electrons from positrons at higher energies and to increase statistics in all channels. Future results from PAMELA and especially AMS-02 on leptons but also all nuclei fluxes will be of great help in feeding the cosmic ray propagation models. Then, the indirect searches for dark matter through charged channels can go on, in particular by looking for fine structures in the spectra. It will then be very interesting (and challenging!) to try to interpret future data and weight them against LHC and direct detection results. Whatever the nature of the source, we might be witnessing the first direct observation of a nearby source of cosmic rays with energies in the range of GeV to TeV. These are exciting times and one might have to wait a little more for the final solution to this cosmic puzzle. The answer will certainly come from a convergence of different messengers information. Thanks to its large field of view, the Fermi gamma-ray telescope will hopefully say something about a nearby source, should it be a pulsar or a more exotic one. Eventually, future large observatories in neutrino and gamma rays (km3net, CTA) will certainly offer the great opportunity to have a deep look into this brainteaser.

\section*{acknowledgement}

I wish to very much thank colleagues who these results have been obtained with: Timur Delahaye, J\"urg Diemand, Fiorenza Donato, David Maurin, Stefano Profumo and Pierre Salati.

\bibliographystyle{aipproc}   

\bibliography{sample}

\IfFileExists{\jobname.bbl}{}
 {\typeout{}
  \typeout{******************************************}
  \typeout{** Please run "bibtex \jobname" to optain}
  \typeout{** the bibliography and then re-run LaTeX}
  \typeout{** twice to fix the references!}
  \typeout{******************************************}
  \typeout{}
 }





\end{document}